\documentstyle[preprint,aps,12pt]{revtex}
\begin{document}
\draft
\hfill\vbox{\baselineskip14pt
             \hbox{\bf PC-16/ISS 2001}
            \hbox{June 2001}}
\baselineskip20pt
\vskip 0.2cm 
\begin{center}
{\Large\bf Temperature Dependent Polarized XANES Spectra for
Zn-doped LSCO system}
\end{center} 
\begin{center}
\large Sher~Alam$^{1}$,~A.T.M.N Islam $^{2}$,~I.~Tanaka$^{2}$,
~P.~Badica$^{3}$, ~H.~Oyanagi$^{1}$,H.~Kawanaka$^{3}$,
~M.O.Rahman$^{4}$~and~T.~Yanagisawa$^{3}$
\end{center}
\begin{center}
$^{1}${\it Photonics, Nat.~Inst.~of~AIST, Tsukuba, Ibaraki 305-8568, Japan}\\
$^{2}${\it Inst.~of~Inorganics Synth.~,~Yamanashi Univ.~Miyame~7,
~Kofu~400-8511}\\
$^{3}${\it Nanoelectronics, Nat.~Inst.~of~AIST, Tsukuba, Ibaraki 
305-8568, Japan}\\
$^{4}${\it  GUAS \& Photon Factory, KEK, Tsukuba, Ibaraki 305, Japan}
\end{center}
\begin{center} 
\large Abstract
\end{center}
\begin{center}
\begin{minipage}{16cm}
\baselineskip=18pt
\noindent
  The cuprates seem to exhibit statistics, dimensionality and phase 
  transitions in novel ways. The nature of excitations [i.e. quasiparticle 
  or collective], spin-charge separation, stripes [static and dynamics], 
  inhomogeneities, psuedogap, effect of impurity dopings [e.g. Zn, Ni] and 
  any other phenomenon in these materials must be consistently understood. 
  Zn-doped LSCO single crystal were grown by TSFZ technique. Temperature 
  dependent Polarized XANES [near edge local structure] spectra were 
  measured at the BL13-B1 [Photon Factory] in the Flourescence mode from 
  10 K to 300 K. Since both stripes and nonmagnetic Zn impurities 
  substituted for Cu give rise to inhomogeneous charge and spin distribution 
  it is interesting to understand the interplay of Zn impurities and stripes.
  To understand these points we have used Zn-doping and some of the results 
  obtained are as follows: The spectra show a strong dependence with respect 
  to the polarization angle, $\theta$, as is evident at any temperature by 
  comparing the spectra where the electric field vector is parallel with 
  ab-plane to the one where it is parallel to the c-axis. By using the 
  XANES [temperature] difference spectra we have determined 
  T* [experimentally we find, T* $\approx$ 160-170 K] for this sample. 
  The XANES difference spectra shows that the changes in XANES features 
  are larger in the ab-plane than the c-axis, this trend is expected 
  since zinc is doped in the ab-plane 
  at the copper site. Our study also complements the results in literature 
  namely that zinc doping does not affect the c-axis transport.

\end{minipage}
\end{center}
\vfill
\baselineskip=20pt
\normalsize
\newpage
\setcounter{page}{2}
\section{Introduction}
	The Cuprates, Nickellates, and Manganites display 
interesting features of spin, charge, lattice and orbital 
orderings. These features lead to the display of a wide 
variety of physical phenomenon, such as Paramagnetism [PM], 
Ferromagnetism [FM], antiferromagnetism [AF], 
Chargetism\footnote{We are coining this terminology since we want 
to emphasize that akin to orderings in magnetism there is a dual 
ordering with respect to charge distribution} [i.e.charge orderings], 
Orbitism [i.e. orbital orderings], Superconductivity, 
``Strange Metal'' [SM] behaviour. In addition to the fundamental 
and interesting physics, these materials offer technological 
use [i.e. Giant Magnetoresistance [GMR] also called Colosssal 
Magnetoresistance [CMR]] and High Temperature Superconductivity [HTSC]
\cite{oya00}. 

	Many HTSC materials are cuprates, which seem to exhibit 
statistics, dimensionality and phase transitions in novel ways. 
The nature of excitations [i.e. quasiparticle or collective], spin-charge 
separation, stripes [static and dynamics], inhomogeneities, psuedogap, 
effect of impurity dopings [e.g. Zn, Ni] and any other phenomenon in 
these materials must be consistently understood. Thus we turn to a
sytematic\footnote{By systematic we mean that we intend to make
using the same crystals all studies, such as XRD, EXAFS, XANES,
IR, Optical, Raman spectroscopies, Specific heat, thermal conductivity,
neutron scattering, NMR and others! } study of LSCO system for a range 
of Zn doping [0.5\%, 1.0 \%, and 2.0 \%]. Here our purpose is to
report on our temperature dependent polarized XANES measurements
on the 1\% LSCO sample. Zn-doped LSCO single crystal were grown by 
TSFZ technique. Temperature dependent Polarized XANES 
[near edge local structure] spectra were measured at the BL13-B1 
[Photon Factory] in the Flourescence mode from 10 K to 300 K. Since 
both stripes and nonmagnetic Zn impurities substituted for Cu give 
rise to inhomogeneous charge and spin distribution it is interesting 
to understand the interplay of Zn impurities and stripes.
To understand these points we have used Zn-doping and some of the results 
obatined are as follows: The spectra show a strong dependence with respect 
to the polarization angle, $\theta$, as is evident at any temperature by 
comparing the spectra where the electric field vector is parallel with 
ab-plane to the one where it is parallel to the c-axis. By using the 
XANES [temperature] difference spectra we have determined 
T*\footnote{Here by T* we simply mean the temperature where there
is a transition in the XANES spectrum. Experimentally we 
find, T* $\approx$ 160-170 K for the current sample.} 
The XANES difference spectra 
shows that the changes in XANES features are 
larger in the ab-plane than the c-axis, this trend is expected since 
zinc is doped in the ab-plane at the copper site. Our study also 
complements the results in literature namely that zinc doping does not 
affect the c-axis transport \cite{uch00}. Local lattice instability
and stripes in the CuO$_{2}$ plane of the La$_{1.85}$ Sr$_{0.15}$ Cu O$_{4}$
using temperature dependent XANES and EXAFS was studied by 
Saini et al. \cite{oya97}. It was concluded in \cite{oya97} that
the temperature-dependent distortions show a maximum around 60 K,
and minimum at 37 K [i.e. T${_c}$] and that two types of doped
charges coexist in different stripes in the superconducting
phase. Here our motivation is a quite different. Among
other details we want to answer in particular the following: Can we 
see a change in the XANES spectrum as a function of temperature, which 
originates from the inhomgeneous charge distribution in cuprates such 
as LSCO. Since as already mentioned above both stripes and nonmagnetic 
Zn impurities substituted for Cu give rise to inhomogeneous charge and spin 
distribution it is interesting to understand the interplay of Zn impurities 
and stripes. Naively we expect that stripes are pinned by Zn and that
leads to suppression of quasiparticle weight. In particular it is
found that the 1/8 hole-concentration anomaly of T$_{c}$, which
is an indication of stripes is induced in Bi$_{2}$Sr$_{2}$CaCu$_{2}$O$_{8}$
by Zn substitution \cite{ako98}. Moreover the angle-resolved
photoemission [ARPES] experiments show that the Zn substitution
suppresses the quasi-particle weight along the (0,0)-$(\pi,\pi)$
direction \cite{whi99}, this suppression is qualitatively same as that of
underdoped La$_{2-x}$ Sr$_{x}$ Cu O$_{4}$ in which vertical stripes 
exist \footnote{These points have also been noted by 
Toyama et al.~\cite{toy00}}. Keeping these remarks in mind a 
little thought tell us that since the 
T$_c$ of 1\%Zn-doped La$_{1.85}$ Sr$_{0.15}$ Cu O$_{4}$
is roughly 20 K, and on the other hand a $T_c \approx 20$ K LSCO system is 
underdoped with T* $\approx$ 160-170 K, we would expect a
T* $\approx$ 160-170 K in a 1\%Zn-doped 
La$_{1.85}$ Sr$_{0.15}$ Cu O$_{4}$ if indeed a relation between
stripes of kind mentioned above exists. Indeed we found a
transition in XANES measured spectrum in the ab-plane as a function
of temperature at 160-170 K, see Fig.~\ref{fig6}. This is very
inspiring since we could experimentally prove the guess about
relationship of stripes and Zn-doping. In some sense we can
map regions of phase diagrams into each other by varying zinc doping
or by increasing stripe effect by going to the underdoped region,
thus probing various regions of the phase diagrams. This also, is in
keeping with our theoretical conjecture, that the 
of superconducting phase is intimately connected with fundamental 
collective excitations [1-dimensional objects i.e. strings]


\section{Experimental Details}
	Conventional traveling solvent floating zone (TSFZ) 
method \cite{tan89} was used for the crystal growth of 
La$_{2-x}$Sr$_{x}$Cu$_{1-y}$Zn$_{y}$O$_{4}$ (x=0.15, y=0.01). 
High purity (99.99\%) powders of La$_{2}$O$_{3}$,~SrCO$_{3}$,~CuO, 
~ZnO were used as raw materials. For feed rod preparation, the raw 
materials were mixed in stoichiometric amount (x=0.15, y=0.01). 
The composition of solvent determined by slow cooling floating 
zone (SCFZ) method \cite{tan99} was LSCZO (80 mol\% (Cu+Zn)O) with 
x=0.4, y=0.013. The crystal was grown in a four mirror type 
infrared heating furnace in oxygen pressure 
$P(O_{2}) =0.2$ MPa at a growth speed of 1 mm/h. The
as-grown crystal was cigar shaped with diameter about 5.5~mm and 
60~mm along the growth direction. It showed metalic lustre and 
confirmed to be single domain by polarizing optical microscopy.

The XANES measurements were conducted at beamline B13-B1 at 
the Photon Factory (PF), Tsukuba, in the Fluorescence mode. 
The electron beam energy was 3.0 GeV and the maximum stored 
current is 400 mA. EXAFS data was collected using fixed-exit 
double crystal Si(111) monochromator. The
first crystal is a water-cooled flat Si(111) monochromator
and the second crystal is sagitally bent to focus the
horizontal beam over $\equiv$ 2 mrad. 
A 19-element solid state detector was used to collect the 
flourescence signal. The large number of detectors allows us 
to cover a sizeable amount of the solid angle of the x-ray 
flourescence emission in addition to giving a high signal 
to noise ratio. The temperatures of sample were controlled
and monitored to within an accuracy of $\pm$ 1 K. 

\section{ Results and Discussion}

	In addition to our experimental work we have generated
theoretically predicted spectra using FEFF8.10 \cite{ank98}
for La$_{2}$Cu O$_{4}$, Fig.~\ref{fig1}, unpolarized 
La$_{1.85}$ Sr$_{0.15}$ Cu$_{0.99}$ 
Zn$_{0.01}$ O$_{4}$, Fig.~\ref{fig2}, and 
La$_{1.85}$ Sr$_{0.15}$ Cu$_{0.99}$ Zn$_{0.01}$ O$_{4}$, 
Fig.~\ref{fig3}, when the polarization vector is along the c-axis.
The definitions of the quantities $\mu$, $\mu_0$ and $\chi$
can be found in \cite{ank98}.
We were not able to generate successfully the XANES
spectra in case of $E||ab$-plane. Let us briefly discuss some simple 
points. By intention we have used the default values in the feff.inp 
file in order to generate the spectra. In order to leave room
for standard comparisons the crystallographic data of 
Radaelli et al., \cite{rad94} was used at 10 K, in our 
FEFF calculations. A rough estimate of La$_{1.85}$ Sr$_{0.15}$ Cu$_{0.99}$ 
Zn$_{0.01}$ O$_{4}$ [Fig.~\ref{fig3}] when the polarization vector is 
along the c-axis [for definition of peaks A please see below],
is  A$_{1}$ $\approx$ 8995 eV, A$_{2}$ $\approx$ 9002  eV, 
A$_{3}$ $\approx$ 9010 eV, thus the energy splittings are
A$_{2}$-A$_{1}$ $\approx$ 7 eV, and A$_{3}$-A$_{2}$ $\approx$ 8 eV.
Even by using almost default input files we can generate
good agreement. We note that the edge energy found by the FEFF8.10 
evaluations is $E_0 =8985.97415 $ eV, and that we have used
the dopant card of atoms 2.50. 

	We have measured the XANES spectrum at temperatures
of 10 K, 15 K, 20 K, 30 K, 40 K, 60 K, 80 K, 100 K, 120 K, 140 K,
160 K, 180 K, 200 K, 220 K, 240 K, 260 K, 280 K, and 300 K
for both polarization directions, i.e. E $||$c-axis and
E$||$ab-plane. Typical raw and subtracted spectra [at 20 K]
are shown respectively in Figs.~\ref{fig4} and ~\ref{fig5}.
The sharpness and detailed nature of the spectra are
clear. Both the raw and subtracted data is extracted with
Ada1. The edge energy used by Ada1 is $E_0=8980.3$ eV. 
Restricting ourselves upto energies of 9050 eV in 
Figs.~\ref{fig4} and ~\ref{fig5}, we can see several
peaks. For the E$||c$-axis case we can see several
peaks A$_{0}$, A$_{1}$, A$_{2}$, A$_{3}$, A$_{4}$, 
and A$_{5}$, similarly for the  E$||$ab-plane
the peaks B$_{0}$, B$_{1}$, B$_{2}$, B$_{3}$ and
B$_{4}$. Peaks A$_{0}$ and B$_{0}$ will not be discussed
here, features A$_{1}$, A$_{2}$, A$_{3}$, B$_{1}$, and
B$_{2}$ are briefly described below.

	As can be seen from our typical measured XANES Spectra, 
the main absorption features are A$_{1}$, A$_{2}$, and A$_{3}$
for the $E||c$ case and B$_{1}$, and B$_{2}$ for $E||ab$.
In large part, the peaks A$_{1}$, and A$_{2}$ are determined
by the multiple scattering of the ejected photoelectrons
off apical oxygen and La and Sr atoms. The peak  B$_{1}$
arises from the multiple scattering off the in-plane oxygen
and copper in the CuO$_{2}$ plane, whereas B$_{2}$ not
only includes the multiple scattering contributions
akin to B$_{1}$ but in addition shows the many body shake-up
satellite of it.

Here we give a rough estimate of energy positions and energy
difference in peaks, leaving more details and analysis for
elsewhere. We find  A$_{1}$ $\approx$ 8994 eV, 
A$_{2}$ $\approx$ 9000 eV, A$_{3}$ $\approx$ 9007 eV,
 B$_{1}$ $\approx$ 9004.5 eV, and  B$_{2}$ $\approx$  9011.5 eV.
Thus the energy splittings between the peaks in the XANES
spectra are as follows: A$_{2}$-A$_{1}$ $\approx$ 6 eV,
 A$_{3}$-A$_{2}$ $\approx$ 7 eV,  B$_{1}$-A$_{2}$ $\approx$ 4.5 eV,
and  B$_{2}$-B$_{1}$ $\approx$ 7 eV. Even with this rough estimate
our values are in good agreement with standard experimental values
quoted for La$_{2}$CuO$_{4}$ and  La$_{1.85}$ Sr$_{0.15}$ Cu O$_{4}$
systems. 

 The temperature dependence of the measured difference 
with respect to [w.r.t] 20 K XANES Spectra  
for La$_{1.85}$ Sr$_{0.15}$ Cu$_{0.99}$ Zn$_{0.01}$ O$_{4}$
for ab-plane and c-axis is shown respectively in Figs.~\ref{fig6}
and ~\ref{fig7} respectively.
The quantity we plotted is XANES difference spectrum.
This is simply defined as the difference between the
XANES spectrum at a temperature T and reference temperature
20 K. The ``difference'' we have plotted is defined to be 
the [approximate] value between the maximum and minimum of the 
difference spectra.
A ``transition'' is clearly seen in
the ab-plane case  at T=T*$\approx 160-170$ K whereas the c-axis data
does not register any measurable
[on this scale] change as is clear from Figs.~\ref{fig6}
and ~\ref{fig7}. This suggest that a ``transition'' in
2D inhomogeneous electron gas. By ``transition'' we mean here
an abrupt change in the XANES spectra given a certain doping and
co-doping [here by Zn] at a certain temperature or temperature
interval. By the pinning of stripes we can probe the strange
metal phase in LSCO material. Theoretically we consider this 
as a charge order-disorder transition. In our scenario of HTSC
theory we consider superconductivity arising from
the dressing of 1-d stripe or string phase. The stripe
arises due to a line of quantum critical points. This
is supported by Monte-Carlo calculations [among other
reasons] of 2D d-p Hubbard model where stripes and
superconductivity exists. The correlation length in HTSC
material is short compared to their conventional cousins,
since here superconductivity arises due to 1-d stripes.
 
\section{Conclusions}
 Temperature dependent Polarized XANES [near edge local structure] 
 spectra were measured at the BL13-B1 [Photon Factory] in the Flourescence 
 mode from 10 K to 300 K for very good quality single crystal of LSCO
 co-doped with 1 \% Zinc at the Cu site. Since both stripes and nonmagnetic 
 Zn impurities substituted for Cu give rise to inhomogeneous charge and spin 
 distribution it is interesting to understand the interplay of Zn impurities 
 and stripes. We recently learned that Pan et al. \cite{pan01} have
 reported microscopic electronic inhomogeneity in the high-Tc 
 superconductor Bi$_{2}$Sr$_{2}$CaCu$_{2}$O$_{8+x}$.
 To understand these and other points we have used Zn-doping and 
 some of the results obtained are as follows: The spectra show a strong 
 dependence with respect to the polarization angle, $\theta$, as is evident 
 at any temperature by comparing the spectra where the electric field vector 
 is parallel with ab-plane to the one where it is parallel to the c-axis. 
 By using the XANES [temperature] difference spectra we have determined 
 T* [experimentally we find, T* $\approx$ 160-170 K] for this sample. 
 The XANES difference spectra shows that the changes in XANES features 
 are larger in 
 the ab-plane than the c-axis, this trend is naively expected since zinc is 
 doped in the ab-plane at the copper site. Our study also complements the 
 results in literature namely that zinc doping does not affect the c-axis 
 transport. We have also generated theoretical spectra from FEF8.10
 which agrees well the experimental results, keeping in mind that
 we have mainly used default FEFF8.10 settings, by intention. However
 we were not able to generate the ab-plane XANES spectra with
 FEFF8.10.

\section*{Acknowledgments}
The Sher Alam's work is supported by the Japan Society for
for Technology [JST].

\begin{figure}
\caption{XANES Spectra calculated using FEFF8.10
for La$_{2}$Cu O$_{4}$}\label{fig1}
\end{figure}
\begin{figure}
\caption{XANES [unpolarized] Spectra for obtained with FEFF8.10
for La$_{1.85}$ Sr$_{0.15}$ Cu$_{0.99}$ Zn$_{0.01}$ O$_{4}$}\label{fig2}
\end{figure}
\begin{figure}
\caption{XANES [c-axis polarized] Spectra for obtained with FEFF8.10
for La$_{1.85}$ Sr$_{0.15}$ Cu$_{0.99}$ Zn$_{0.01}$ O$_{4}$}\label{fig3}
\end{figure}
\begin{figure}
\caption{Measured Raw XANES [ab-plane and c-axis polarized] Spectra for
for La$_{1.85}$ Sr$_{0.15}$ Cu$_{0.99}$ Zn$_{0.01}$ O$_{4}$ at 20 K}
\label{fig4}
\end{figure}
\begin{figure}
\caption{Subtracted [using Ada1] measured XANES [ab-plane and c-axis 
polarized] Spectra for 
La$_{1.85}$ Sr$_{0.15}$ Cu$_{0.99}$ Zn$_{0.01}$ O$_{4}$ at 20 K}
\label{fig5}
\end{figure}
\begin{figure}
\caption{Temperature Dependence of the measured difference [w.r.t 20 K] 
XANES [ab-plane polarized] Spectra 
for La$_{1.85}$ Sr$_{0.15}$ Cu$_{0.99}$ Zn$_{0.01}$ O$_{4}$}\label{fig6}
\end{figure}
\begin{figure}
\caption{Temperature Dependence of the measured difference [w.r.t 20 K] 
XANES [c-axis polarized] Spectra for 
La$_{1.85}$ Sr$_{0.15}$ Cu$_{0.99}$ Zn$_{0.01}$ O$_{4}$}\label{fig7}
\end{figure}
\end{document}